\documentstyle[aps,twocolumn]{revtex}

\begin{document}


\title{Vortices In Density Wave Systems\\
       Subject To Transverse Electric Fields}

\author{Akakii Melikidze}
\address{Physics Department, Princeton University, Princeton, NJ
08544}

\date{February 11, 1998}

\maketitle

\begin{abstract}
In this paper we predict many interesting new properties of vortices in
highly anisotropic density wave systems subject to strong
transverse electric fields. We mainly concentrate on
ground state properties. Besides electric field-induced vortices
we consider also thermally activated vortices. A new type of
temperature-driven transition between two different phases
of density waves in strong fields is predicted and several
new properties of those phases are reported.
\end{abstract}



\section{Introduction}

It is now well realized that the dynamics of topological defects
plays an important role in low dimensional electronic systems. In
a particular case of density wave (DW) ground states (for a
review see \cite{Gruner}) it was found that topological defects
are responsible for such intersting phenomena as generation
of narrow-band noise \cite{Ong}, non-linear current-voltage
characteristics in strong fields (the phenomenon associated with
the so-called ''phase slippage'')
\cite{Ramakrishna,Duan}. More recently, quantum phase slip
through creation of vortices was proposed \cite{Maki} to explain
low-temperature properties of spin-density wave compound
$(TMTSF)_2PF_6$ \cite{Mihaly,Nad}.

This paper was originally initiated by the prediction
\cite{Hayashi} that DW systems should possess a novel phase in strong
electric fields applied transverse to the direction of highest
conductivity. This novel phase was called ''mixed state'' in
analogy with the mixed state of type-II superconductors, and was
characterized by the presense of vortices. We have undertaken an
extensive study of the predicted transition. The possibility of
transverse electric field-induced transition to metallic state
(type-I-like) was also investigated and a temperature-driven
transition between type-I and type-II regions of the phase
diagram was predcited. We report these findings here as well as
some interesting anticipated properties of the above-mentioned
phases such as, e.g., non-exponential screening of electric fields by
DW systems in ''mixed state''.


\section{Ginsburg-Landau Theory}

We adopt macroscopic mean-field description of the dynamics of DW
systems and make use of (time-independent) Ginsburg-Landau free energy
\cite{Gruner}
which is a functional of complex order parameter
$\Delta({\bf r})=|\Delta({\bf r})|\exp(i\phi)$,
absolute value of which is DW gap in the
electronic spectrum. For the case of highly
anisotropic system with an open Fermi surface:
${\bf k}=(\pm k_F(1+\gamma\cos(bk_y)),k_y,k_z)$
(here $x$ is the direction of highest conductivity - usually the
direction along conducting chains,
$y$ is the direction of strongest Fermi surface wrapping,
$\gamma$ is the anisotropy parameter, $b$ is
lattice constant along $y$; we neglect dispersion along $z$)
this functional is \cite{coeff_1}:
\begin{eqnarray}
F_{inhom} &=& \int d{\bf r}\frac{K}{2|\Delta_0|^2}\bigg[
            |\partial_x\Delta|^2+\gamma^2|\partial_y\Delta|^2\bigg]
\label{action1}
\end{eqnarray}
We would like to emphasize the fact that the two terms
in $F_{inhom}$ are of totally different origin.
As is well known \cite{Gruner}, in the system described above
a density modulation appears with the wavevector:
${\bf q}={\bf q}_0=(2k_F,\pi/b,0)$: $\rho_{spin}\propto \Delta \cos({\bf 
q_0r})$.
Spatial variation of $\Delta$ gives rise to the effective
shift in ${\bf q}$ by $\delta{\bf q}=-i\vec{\nabla}\log
\Delta$. Thus the gap is developped at the wave-vector which is slightly
off its optimal value ${\mathbf q}_0$.
This costs extra energy: $\delta E \propto |\delta
\mathbf{q}|$ - it is linear with $\delta{\mathbf q}$ which is
quite costly for small deviations.
Instead the system rearranges its {\it electron
density} in such a way that {\it locally} $2k_F$ coincides with the new
DW wave-vector. There is no more linear contribution to the
energy, however another contribution arises - the energy of the
non-homogenious electron density distribution. But the fact is
that this density (per chain) is:
\begin{eqnarray}
\rho = \frac{1}{\pi}(-i\partial_x)\log\Delta
\label{induced_density}
\end{eqnarray}
and the extra energy is: $\delta E \propto \rho^2 \propto (\delta
q_x)^2$ - it
is {\it quadratic} in $\delta q_x$ and therefore wins for small
distorions of DW gap. This phenomenon is usually described as
follows: ''particle density follows the variations of the gap''.
At the same time the gradient in 
y-direction does not lead to any change in the local charge density; its 
contribution to the free energy is of purely elastic character.

In this paper we shall be interested in the response of DW system
to external {\it electric} fields.
We take account of the effect of
electric field by introduction of the
following term into the free energy functional \cite{Hayashi}:
\begin{eqnarray}
F_{el} =\int d{\bf r}\bigg[
Je\rho\varphi - 
\frac{1}{8\pi} \bigg(\epsilon |\vec{\nabla} \varphi|^2 + 
\lambda_{qp}^{-2}\varphi^2\bigg) + \varphi \rho^{ext} \bigg]
\label{action2}
\end{eqnarray}
Here the first term describes coupling of electric field to the charge
induced by DW distortion, the term in parenthesis represents
energy associated with electric field itself and with the
coupling of electric field to free charge carriers escited over
the DW-gap. The last term represents the interation of electric
field with external charges \cite{coeff_2}.
Strictly speaking Eqn.(\ref{action2}) is the {\it action} of the
system taken with minus sign. However we shall
use the term ''free energy funciotnal'' to denote a functional of
which an extremum determines the ground state.
We also emphasize that electric field 
couples only to x-component of the gradient of the gap - only this 
component induces electric charge!

Important point here is that our action is essentially different
from that of ref.\cite{Hayashi} in that it contains wave
vector-dependent dielectric constant $\epsilon=\epsilon(k)$. It
appears because the ground state of DW, in complete analogy with the
low-$T$ ground state of {\it semiconductors}, is highly polarizable due
to excitations of {\it virtual} electron-hole pairs. This should not be
mixed neither with the screening due to thermally excited
quasiparticles,
nor should it be mixed with the screening due to spatial
distortion of the order parameter(see below).
We stress that this dielectric function is {\it generic} for a
gapped electronic system.
It can be calculated using simple anisotropic
two-band semiconductor model \cite{Gruner}:
\begin{eqnarray}
\epsilon(k) \sim \frac{\lambda_{TF}^{-2}}{k^2 + \xi(k)^{-2}}
\label{epsilon}
\end{eqnarray}
Here $\lambda_{TF}$ is Tomas-Fermi screening length in th
metallic state, $\xi(k)$ is anisotropic coherence length in DW
state. (The value $\epsilon(0)$ is usually huge: for instance
$\epsilon(0) \sim 10^5$ for $(TMTSF)_2PF_6$). Now we turn to the
analysis of the response of DW systems to external electric fields.


\section{Weak Fields}

In the absense of external fields the phase of the order parameter
$\Delta=|\Delta|\exp(i\theta)$ is the so-called degeneracy parameter:
the free energy is globally $U(1)$ invariant - it does 
not change 
if we make a global substitution $\theta \rightarrow \theta + const$. 
Consequently there is a Goldstone (the so-called {\it phason}) gapless mode 
associated with this degeneracy. At the same time the orthogonal
{\it amplitudon}
mode, in which $|\Delta|$ is varied, has a gap in the spectrum at $k=0$.
Therefore we expect that in the limit of weak fields the response of the 
system is given by the excitation of phason mode only. Thus we set
$|\Delta|$ to be constant.
After Fourier transform:
\begin{eqnarray}
F_{inhom}+F_{el} &=& \int d{\bf r}\bigg[
\frac{Kk_\gamma^2}{2}|\theta_k|^2 +
Je\varphi_kik_x\theta_{-k}\nonumber \\
&-&\frac{1}{8\pi}(\epsilon k^2+\lambda_{qp}^{-2})|\varphi_k|^2 
+\varphi_k\rho_{-k}^{ext} \bigg]\nonumber \\
&=& \int d{\bf r}\bigg[
\frac{Kk_\gamma^2}{2}|\theta_k -
i2ek_x\varphi_k/v_Fk_\gamma^2|^2 \nonumber \\
&-& \frac{1}{8\pi}(\epsilon k^2+\Lambda_k^{-2})|\varphi_k|^2
+\varphi_k\rho_{-k}^{ext} \bigg]\nonumber \\
\label{action3}
\Lambda_k^{-2} &=& \lambda_{qp}^{-2} + 
f_s\lambda_{TF}^{-2}\frac{k_x^2}{k_\gamma^2}
\label{screen}
\end{eqnarray} 
the minimization with respect to $\theta_k$ becomes trivial. Then
the remaining 
effective action for the electric field leads to an anisotropic screening 
length $\lambda$. Screening length in y-direction is 
$\lambda_y=\sqrt{\epsilon(\lambda_{qp}^{-1})}\lambda_{qp}$ 
and is due to interaction of electric field with quasiparticles excited 
over the gap. On the other hand screening in x-direction is now 
determined by both quasiparticles and phasons leading to screening length
$\lambda_x=\lambda_{TF}$ \cite{pinning}.
This is again a manifestaion of the fact that only the 
phason mode along x-direction couples to electric field
\cite{omit}.


\section{Strong Fields}

As we have just seen, electric fields in the $x$-direction (along
the chains) are screened at the short lengths $\sim \lambda_{TF}$
at all temperatures. In contrast to that, $E_y$ component of the
field can penetrate far enough into the bulk since its screening
is determined be thermally excited quasiparticles the
concentration of which is exponentially decreased at low
temperatures. However in strong electric fields the situation
changes: $\Delta$ can no longer be considered 
constant and it may be energetically 
favourable for the system to expell electric field from the sample by 
spatial variation of the magnitude of the gap \cite{variation}.
Firstly, it's obvious that in the limit $E_y\rightarrow\infty$ the state of 
the system is metallic. A simple argument for this is that in 
competition between condensate energy
$-n(\epsilon_F)\Delta^2/2$
(here $n(\epsilon_F)=N_{\perp}/\pi v_F$ is density of states at Fermi 
   level, $N_\perp$ is density of chains in the plane perpendicular to the 
   chains)
and electrostatic energy $ED/8\pi$ the 
latter always wins in the limit of strong fields: the gap vanishes giving 
rise to screening. Thus in the region of strong fields there is a thin 
layer near the 
edge of the sample in which the state is metallic. 
Electric field in this layer is screened at the distance $\sim \lambda_{TF}$.
However, taking into account that actually the continuous model used here 
is not aplicable at such small distances, one should substitute
$\lambda_{TF}$ for interchain distance. This corresponds to a 
metallic layer with the thickness of as small as a single interchain distance.
However, in the bulk the electric field is absent and DW state
is restored.

Thus we establish that there must exist a critical value of $E_y$
for which a transition occurs from DW to some other phase. 
As the field is increased from small values two possibilities may occur:
$\Delta$ can jump to a smooth configuration (Type-I) or a
configuration with topological defects \cite{Hayashi} (Type-II). The latter
means that the ground state is characterized by the presense of
vortices - centers the in-plane circulation of the phase of the
order parameter around which is non-zero.
These two 
possibilities correspond to the two types of superconductors in 
magnetic field. An ordinary isotropic superconductor can be distinguished
between these two types by the evaluation of Ginsburg-Landau parameter
$\kappa=\lambda/\xi$. Values $\kappa<1/\sqrt{2}$ and $\kappa>1/\sqrt{2}$
correspond to type-I and type-II respectively. In our case this simlpe 
argument is not applicable since DW materials are usually highly anisotropic. 
Indeed, usually
along $x$-direction one has $\lambda_x \ll \xi_x$ whereas
along $y$-direction either of $\lambda_y \ll \xi_y$ and $\lambda_y \gg
\xi_y$ can hold depending on temperature: $\lambda_y$
exponentially increases with decreasing temperature (see below
for a more detailed discussion).
Instead we shall use a more physical
argument to decide between Type-I and Type-II.
Namely, we shall evaluate the critical field for
these two cases. 
 Then we can argue that the system 
will make a transition to the state for which the critical field 
is lower.


\subsection{Type I}

 The critical field for this case can 
be obtained by equating condensate and electrostatic free energy densities:
\begin{eqnarray}
\frac{ED}{4\pi} &=& \frac{1}{2}n(\epsilon_F)\Delta_0^2\nonumber \\
D_c^I &=& \frac{\Delta_0}{2e\lambda_{TF}}
\sim \frac{d_0}{\xi_x\lambda_{TF}}
\label{Ec}
\end{eqnarray}
Here we have introduced $d_0=v_F/2e$ - a parameter with the
dimensionality of electric dipole moment (see below).
N.B. We use the value of the dielectric constant $\epsilon \sim
1$ because the metallic breakdown  will occur in a thin
layer near the boundary with the thickness $\sim \lambda_{TF} \ll
\xi_y$ - polarization will not develop at such small scales!


\subsection{Type II - Vortices}

First of all let us
clearify what is a vortex in a DW system. DW state is a ground
state with broken $U(1)$ global invariance. This means that the
order parameter is a complex number: $\Delta =
|\Delta|\exp(i\theta)$. The order parameter, as a function of
coordinates, should be univalued. This implies that, as we travel
along a closed contour, the phase of the order parameter can get
an increment $2n\pi$ where $n$ is integer called {\it winding
number}. The corresponding
texture of the order parameter is called {\it n-times quantized
vortex}. Let us now consider a singly-quantized vortex located at
$\vec{r}=0$:
\begin{eqnarray}
\Delta(\vec{r}) = |\Delta(r)|\exp(i\phi)
\label{texture}
\end{eqnarray}
where $\phi$ is asimuthal angle in the $x$-$y$ plane.
The absolute value of the gap is assumed
to be constant everywhere except regions of the size $\xi$
(correlation  length) near the vortex centers.
This assumtion constitutes the
so-called London approximation. From Eqn.(\ref{texture}) we can
calculate induced charge density (per chain) using
Eqn.(\ref{induced_density}):
\begin{eqnarray}
e\rho = \frac{e}{\pi}\frac{\sin \phi}{r}
\label{induced_charge_density}
\end{eqnarray}
We see that it falls off as $1/r$ - it is highly non-local! However
the {\it screening} makes the fall-off {\it
exponential}. To see how this happens we make the following
substitution in Eqn.(\ref{action3}) in order to take vortex
degree of freedom into account \cite{Hayashi}:
$(\vec{\nabla} \theta)_k \rightarrow i\vec{k}\theta_k^{phason}+
(i\vec{k}\times \hat{z}/k^2)n_k$ where
$n(r)=2\pi \sum n_i \delta(r-r_i)$
is vorticity and $n_i$ is the winding number of a vortex at $r_i$. 
Minimizing the free energy functional with respect to $\theta_k^{phason}$
we get an effective action:
\begin{eqnarray}
F_{inhom}+F_{el} &=& \int \frac{d^2\vec{k}}{(2\pi)^2} \bigg[
\frac{K\gamma^2}{2k_\gamma^2}\bigg|n_k-\frac{ieJk_y}{K}\varphi_k\bigg|^2
\nonumber \\
&-& \frac{1}{8\pi}(\epsilon k^2+\lambda_{tf}^{-2})|\varphi_k|^2 
+\varphi_k\rho_{-k}^{ext} \bigg]
\label{Feff}
\end{eqnarray}
Electric potential produced by vortices is obtained by the
minimization of Eqn.(\ref{Feff}) with respect to $\varphi_k$ and
is given by:
\begin{eqnarray}
\varphi_k = \frac{4\pi ie\gamma^2 J}{\epsilon k^2+\Lambda_k^{-2}}
\frac{k_y}{k_{\gamma}^2}n_k
\label{Potential}
\end{eqnarray}


\subsection{Single Vortex}
Now $n(r)=2\pi\delta(r)$ and the
electrostatic potential produced by a single vortex is given by the
Fourier transform of Eqn.(\ref{Potential}). It is hard to obtain an
analytical expression in the general case, but one can get an
idea of what this potential looks like from the consideration of
the purely isotropic case: $\gamma=1$, $\Lambda_k=\Lambda=const$.
With this simplification the potential is given by:
\begin{eqnarray}
\phi(\vec{r}) &=& - 2e\gamma^2J \sin{\theta} f(r)\nonumber \\
f(r) &=& \int dk \frac{J_1(kr)}{\epsilon
k^2+\Lambda^{-2}} \\
&=& \left\{
\begin{array}{ll}
\frac{r}{2\epsilon}\log{\frac{\epsilon^{1/2}\Lambda}{r}}
    &r\ll \epsilon^{1/2}\Lambda,\nonumber \\
\frac{\Lambda^2}{r}, &r\gg \epsilon^{1/2}\Lambda.
\end{array}
\right.
\label{PotentialofR}
\end{eqnarray}
Here $J_1$ is first Bessel function.
The effect of huge actual anisotropy in $\Lambda$ can be taken
into account qualitatively in the following way. First of all we
may notice that there will be still two characteristic regions: $r
\ll \epsilon^{1/2}\Lambda(\theta)$ and $r \gg \epsilon^{1/2} 
\Lambda(\theta)$ where
$\Lambda(\theta)$ is now angle-dependent crossover scale (which
is, of course, extremely anisotropic).
Then we may also notice that in the large-$r$ region $f(r)$ is
still angle-independent: it is given by
$f(r) \approx \Lambda_{av}^2/r$ where $\Lambda_{av}$ is some
''average'' screening length. Important point here is that the
{\it total charge density} in the large-$r$ region falls off
{\it exponentially}: one can check that by taking laplacian of
$\varphi(\vec{r})$. But the above mentioned extreme
anisotropy comes to play in the small-$r$ region - the core of
a vortex is highly anisotropic and hard to analyze.

Some of the vortex properties, however, allow exact description. From
Eqn.(\ref{Potential}) it can be inferred that the total charge
of a vortex iz zero, but there is a non-zero electric dipole
moment directed along $y$:
\begin{eqnarray}
d_y&=&2\pi \epsilon eJ\lambda_{qp}^2=d_0\frac{\epsilon
f_s}{2(1-f_s)}
\label{Dipole}
\end{eqnarray}
Here $d_0=v_F/2e$.
$d_y$ is exponentially increased as $T\rightarrow 0$. It should be
noted that this expression was derived for a vortex in the bulk;
it is expected that for vortices near boundary (the case which is
relevant for vortices produced by external fields - see below)
the induced dipole moment is reduced.


\subsection{Critical Field}
Now we estimate the critical field at which
an appearence of a single vortex becomes energetically
favourable. In order to do that we substitute
Eqn.(\ref{Potential}) (with $n_k=2\pi$) back into
Eqn.(\ref{Feff}):
\begin{eqnarray}
F_{eff} &=& \int d^2\vec{k}
            \frac{K\gamma^2}{2k_\gamma^2}
            \bigg(\frac{\epsilon k^2+\lambda_{TF}^{-2}}
            {\epsilon k^2+\Lambda_k^{-2}}\bigg)
            - d_y E_y^{ext} \\
        &=& \frac{\pi K \gamma}{2} \frac{\lambda_{qp}}{\lambda_{TF}}
            \log{\frac{W}{\epsilon^{1/2}\lambda_{qp}}} -
            d_y E_y^{ext}
\label{Change}
\end{eqnarray}
Here $W$ is the size of the system in the y-direction.
From Eqn.(\ref{Change}) and Eqn.(\ref{Dipole}) we obtain the critical field:
\begin{eqnarray}
D_c^{II} &=& \frac{ \gamma d_0}{4\lambda_{qp}\lambda_{TF}}
\log{\frac{W}{\epsilon^{1/2}\lambda_{qp}}}
\label{Dc2}
\end{eqnarray}
Comparing this expression with its superconducting analog one may call
$d_0$ an {\it ''electric dipole moment quantum''}.
The dependence of the critical field on temperature is given through:
$\lambda_{qp}^{-2}=\lambda_{TF}^{-2}\sqrt{2\pi \Delta_0/T} \exp(-\Delta_0/T)$.
So the critical field for vortex mixed state can be significantly
lowered by increasing the screening length in the y-direction as
$T \rightarrow 0$. As we do so, an interesting situation can
occur: $D_c^{II}$ can be made smaller than $D_c^{I}$
signalling a {\em temperature-driven transition between Type-I
and Type-II-like ground states}! A more intuitive explanation of
this transition is obvious: one can, in principle, evaluate
$\kappa_x = \lambda_x / \xi_x$ and $\kappa_y = \lambda_y /
\xi_y$. Usually one has
$\kappa_x \ll 1$ while $\kappa_y$
is exponentially temperature dependent. One can argue then that the type
of the ground state is determined by the geometric mean of
the two kappas: $\kappa = \sqrt{\kappa_x\kappa_y}$. This ''mean''
Ginzburg-Landau parameter $\kappa$ is also temperature-dependent. This
raises a possibility of {\it Type-I/Type-II temperature-driven 
transition} as some
critical value $\kappa_c \sim 1$ is passed by $\kappa$ in a temperature
sweep. 
To our knowledge this is the first time that such type of
transition is predicted.


\subsection{Screening Properties}
From the physical point of view the
reason why the appearence of vortices becomes favourable at high
fields is that vortices can screen external electric fields. In
order to show this we first introduce a ''dense limit
approximation'' \cite{Hayashi} in which the everage distance between
vortices is
assumed to be much less than the characteristic length of
the external electric field variation. Then one
can take the vorticity $n(\vec{r})$ to be a continuous
function rather than the sum of delta-functions. The
minimization of the effective energy Eqn.(\ref{Feff}) with respect
to $n_k$ becomes trivial:
$n_k = d_0^{-1}E_k^y$
and leaves us with the effective free energy which decribes the screening
with the screening length $\lambda=\lambda_{TF}$. In fact this
contradicts to the above made ''dense limit approximation'',
nevertheless the conclusion about screening remains valid if we
assume that the screening length is rather equal to the average inter-vortex
distance: $\lambda=n^{-1/2}$. But now the screening length
becomes coordinate dependent leading to a {\it non-exponential
screening}. Indeed, Poisson eqaution $\varphi''-\lambda^{-2}\varphi=0$
in this case reads $\varphi''+d_0^{-1} \varphi \varphi' = 0$.
Its solution is:
\begin{eqnarray}
\varphi(y) = \frac{2d_0}{y+2d_0\varphi(0)^{-1}}
\label{phi}
\end{eqnarray}
At distances $\sim \lambda_{qp}$, however, the exponential screening due to
quasiparticles comes to play. So in the bulk the DW state is
restored. Thus the electric field-induced transition to a vortex
state is a {\it surface effect}. Therefore it is unlikely
to be detected in any transport measurements
\cite{spectral_flow}.
On the other hand
the best way to observe such a transition would be to measure
directly the screening length. It would be expected that the
screening length be $\sim \lambda_{qp}$ for field below critical
whereas for fields above critical it 
collapses to some small value. Such experiments are currently
underway \cite{Melikidze}.


\section{Thermally Excited Vortices}

An alternative to the production of vortices in DW systems by
applying transverse electric field at low temperature is their
thermal activation in the temperature region near $T_c$.
We have already calculated the energy
required to produce a singly-quantized vortex in the absense of
external electric field (Eqn.\ref{Change}) - this energy is given
by \cite{unit}:
\begin{eqnarray}
    E_a  =  \frac{\pi K \gamma}{2} \frac{\lambda_{qp}}{\lambda_{TF}}
            \log{\frac{W}{\epsilon^{1/2}\lambda_{qp}}}
\label{activation}
\end{eqnarray}
At low temperature this energy is big, but as we get closer to
the transition point $E_a$ decreases rapidly because of
decreasing rigidity $K$ (it is proportional to the condensate
density). Moreover, $K$ is renormalized to smaller values in the 
vicinity of
critical point by the fluctuations of the order parameter.
The description of the dynamics of thermally excited vortices in
external transverse electric fields is beyond the scope of this article and
will be reported elsewhere; however a simple picture can be
outlined here. The key point is that there is a gas of
thermally activated vortices at some finite temperature.
We shall restrict ourselves to only two kinds of vortices, namely
those with winding numbers $n_i=\pm 1$ and dipole momenta
$d_y = \pm |d_y|$ respectively. It is assumed that the
excitation of vortices with higher winding numbers is suppressed
be their bigger energies. Then applying transverse
electric field one would be able to polarize vortex gas.
The dielectric constant of the vortex gas in the case of {\it weak
fields} is:
\begin{eqnarray}
\epsilon = \frac{4\pi n d_y^2}{T}
\label{epsilon_v}
\end{eqnarray}
Here $n$ is the density of vortices. $n$ also enters the
expression for the correlation length in the gas of vortex dipoles.
Thus one would be able to extract $n$ from experiments concerned
with the response of DW materials to transverse electric fields.
Such experiments are currently underway \cite{Melikidze}.


\section{Conclusions}

We have performed an extensive analysis of vortices in highly
anisotropic quasi-1D DW systems. This work builds a more
detailed and correct picture of the structure and dynamics of vortices as
compared to \cite{Hayashi} and also predicts several
unique properties of such vortices never encountered before with
the case of their superconducting counterparts. Among those are
the possibility of a {\it temperature-controlled phase transition
between Type-I and Type-II-like phases}; the {\it non-exponential
screening of strong electric fields by vortices at comparebly large scales} as
opposed to the incorrect conclusion about short range exponential
screening drawn in \cite{Hayashi}; {\it temperature controlled ''flux
quantum'' of a vortex - its dipole moment}. At the end we would like
to point out that the exploration of the dynamics of topological
defects in DW systems is a relatively new and actively developing
area of condensed matter physics with many surprises and,
possibly, some connections to other branches of physics (see e.g.
a recent paper \cite{Hayashi_new} and references therein).


\section*{Aknowledgements}

We would like to thank S. Sondhi and O. Motrunich for stimulating
discussions and M. V. Feigelman for pointing out some of the
relevant bibliography. We are also grateful to I. Nemenman and O. Motrunich
for careful reading of this article and making numerous usefull
comments. We would like to especially aknowledge numerous
insightful comments on this work by D. Huse and P. M. Chaikin.




\begin{references}
\bibitem{coeff_1}
   The coefficients in the equation are given as follows
   \cite{Fukuyama}:
   $K=N_{\perp}v_Ff_s/2\pi$, $N_\perp$ is density of chains in the plane
   perpendicular to chains,
   $f_s(T)\approx7\zeta(3)|\Delta|^2/4\pi^2T_c^2$ for $T\approx T_c$ and
   $f_s(T)\approx1-\sqrt{2\pi\Delta_0/T}exp(-\Delta_0/T)$ for $T\ll T_c$
   is the dimmensionless condensate density $\rho_s/\rho_0$, $\Delta_0=1.7T_c$
   is the gap at $T=0$;
   $\gamma^2=\xi_y^2/\xi_x^2$ is the anisotropy parameter,
   $\xi_x$ and $\xi_y$ are correleation
   lengths in $x$ and $y$ directions respectively.
\bibitem{coeff_2}
   Here $J=N_{\perp}f_s/\pi$; $\lambda_{qp}^{-2}=(1-f_s)\lambda_{TF}^{-2}$ is
   inverse squared screening length due to quasiparticles excited over 
   the gap, $\lambda_{TF}^{-2}=8N_\perp e^2/v_F$ is inverse squared Tomas-Fermi 
   screening length in the metallic state \cite{Hayashi}. See main text for the
   discussion of $\epsilon$.
\bibitem{omit}
   The polarization contribution to the screening in the
   $x$-direction is omitted because $\epsilon(\lambda_{TF}^{-1})
   \approx 1$.
\bibitem{unit}
   This is the energy of a vortex per unit length.
   For a vortex in a single plane one should multilpy this
   expression by interplane distance.
\end{references}
\end{document}